\documentstyle[twoside,epsf,fleqn,espcrc2]{article}
\gdef\etal{{\it et al.}}
\title{Proposal to search for a monochromatic component of\\
solar axions using $^{57}$Fe
\thanks{Proceedings of the Axion Workshop, University of Florida,
Gainesville, Florida, USA, 13--15 March 1998, edited by P.~Sikivie,
to be published in Nucl. Phys. B Proc. Suppl.}}

\author{Shigetaka~Moriyama
\address{Department of Physics,
School of Science, University of Tokyo,\\
7-3-1 Hongo, Bunkyo-ku, Tokyo 113-0033, Japan}}

\begin{document}

\begin{abstract}
An experimental method is introduced
to search for almost monochromatic solar axions.
In this method, we can search for solar axions
by exploiting the axion-quark coupling,
not relying on the axion-photon coupling at all.
A recent experimental result of Kr\v cmar is presented.
\end{abstract}

\maketitle

\section{Introduction}
The most attractive solution of the strong $CP$ problem
is to introduce the Peccei-Quinn global symmetry
which is spontaneously broken at energy scale $f_a$ \cite{PQ77}.
Although the original axion model has been experimentally excluded,
variant ``invisible'' axion models are still viable.
Such models are referred to as hadronic \cite{Kim} 
and Dine-Fischler-Srednicki-Zhitnitski\v\i\ \cite{DFSZ} axions.
At present, these ``invisible'' axions are constrained by
laboratory searches and by astrophysical and cosmological arguments.
The well-known window for $f_a$
which escapes all the phenomenological
constraints is $10^{10}$--$10^{12}$\,GeV.
This window is frequently quoted
because it is cosmologically interesting.

Besides this, there is another window
around $10^6$\,GeV for the hadronic axions.
This is usually called the hadronic axion window.
A careful study \cite{Chang93} of the window
revealed that $f_a$ in the range $3\times 10^5$--$3\times 10^6$\,GeV
cannot be excluded by the existing arguments,
because most of them were based on the axion-photon coupling
which is the least known parameter 
among those describing the low energy dynamics of the hadronic axions.
In addition, it is argued that $f_a$
around $10^6$\,GeV is desirable to explain
the $ee\gamma\gamma+/\!\!\!\! E_T$ event
in the CDF experiment \cite{CDF}.
Furthermore, a recent paper indicates that axion might
be a hot dark matter if $f_a \sim 10^6$\,GeV \cite{Murayama}.

Although several authors \cite{SikPas,Bibber91}
proposed experiments to search for the axions
with $f_a$ around $10^6$\,GeV,
all of the experiments clearly rely on the axion-photon coupling
both at the source and at the detector.
The methods exploit the Primakoff effect;
photons in the Sun are converted into axions,
which are usually called the solar axions,
and they are re-converted into X rays in a laboratory.
The experiments to search for the solar axions
by the Tokyo group \cite{Sumico},
by the SOLAX collaboration \cite{SOLAX},
and by a pioneer \cite{Lazarus}
rely on the axion-photon coupling.
The experiment \cite{Ressell},
in which an emission line arising from the radiative decay
of axions in the halo of our Galaxy is searched for,
also proceeded through the axion-photon coupling.
Thus there have been no experimental alternatives
to test the hadronic axion window
without relying on the axion-photon coupling.

This encouraged me to invent a new method
to detect axions by exploiting axion-quark couplings,
but not by relying on the axion-photon coupling.

\section{$^{57}$Fe in the Sun}
Because of the axion coupling to nucleons,
there is another component of solar axions.
If some nuclides in the Sun have $M1$ transitions
and are excited thermally,
axion emission from nuclear deexcitation could be also possible.
$^{57}$Fe can be a suitable axion emitter
for the following reasons:
(i) $^{57}$Fe has an $M1$ transition
between the first excited state and the ground state,
(ii) the first excitation energy of $^{57}$Fe
is 14.4\,keV, which is not too high compared with the temperature
in the center of the Sun $(\sim 1.3\,\rm keV)$,
and (iii) $^{57}$Fe is one of the stable isotopes of iron,
which is exceptionally abundant among heavy elements in the Sun.
The natural abundance of $^{57}$Fe is 2.2\%.
If the axion exists, therefore, strong emission
of axions is expected from this nuclide.

\begin{figure}
  \begin{center}
    \leavevmode
    \epsfxsize=6.4cm
    \epsffile{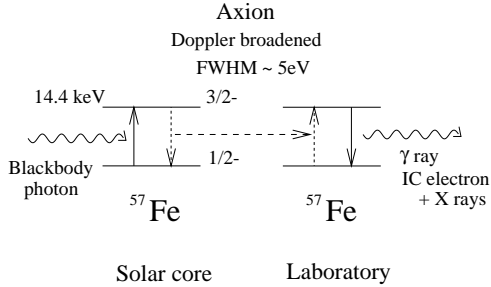}
  \end{center}
  \caption{Principle of the proposed experiment.}
  \label{fig:principle}
\end{figure}

These monochromatic axions would excite the same kind of nuclide
in a laboratory because the axions are Doppler broadened
due to thermal motion of the axion emitter in the Sun,
and thus some axions have energy suitable to excite the nuclide.
Fig.\ \ref{fig:principle} shows
the principle of the proposed experiment.

I propose to search for the axions by detecting this excitation.
Since both the emission and absorption occurs
via the axion-nucleon coupling
but not via the axion-photon coupling,
this method is free from the uncertainty
of the axion-photon coupling.
In addition, this method has merits that
there is no need to tune the detector to a mass of the axions
and that the sensitive mass region can be much
higher than that of the proposed experiment \cite{Bibber91},
in which it is restricted by high pressure of buffer gas.

\section{Expected event rate}
\begin{figure}
  \begin{center}
    \leavevmode
    \epsfxsize=6cm
    \epsffile{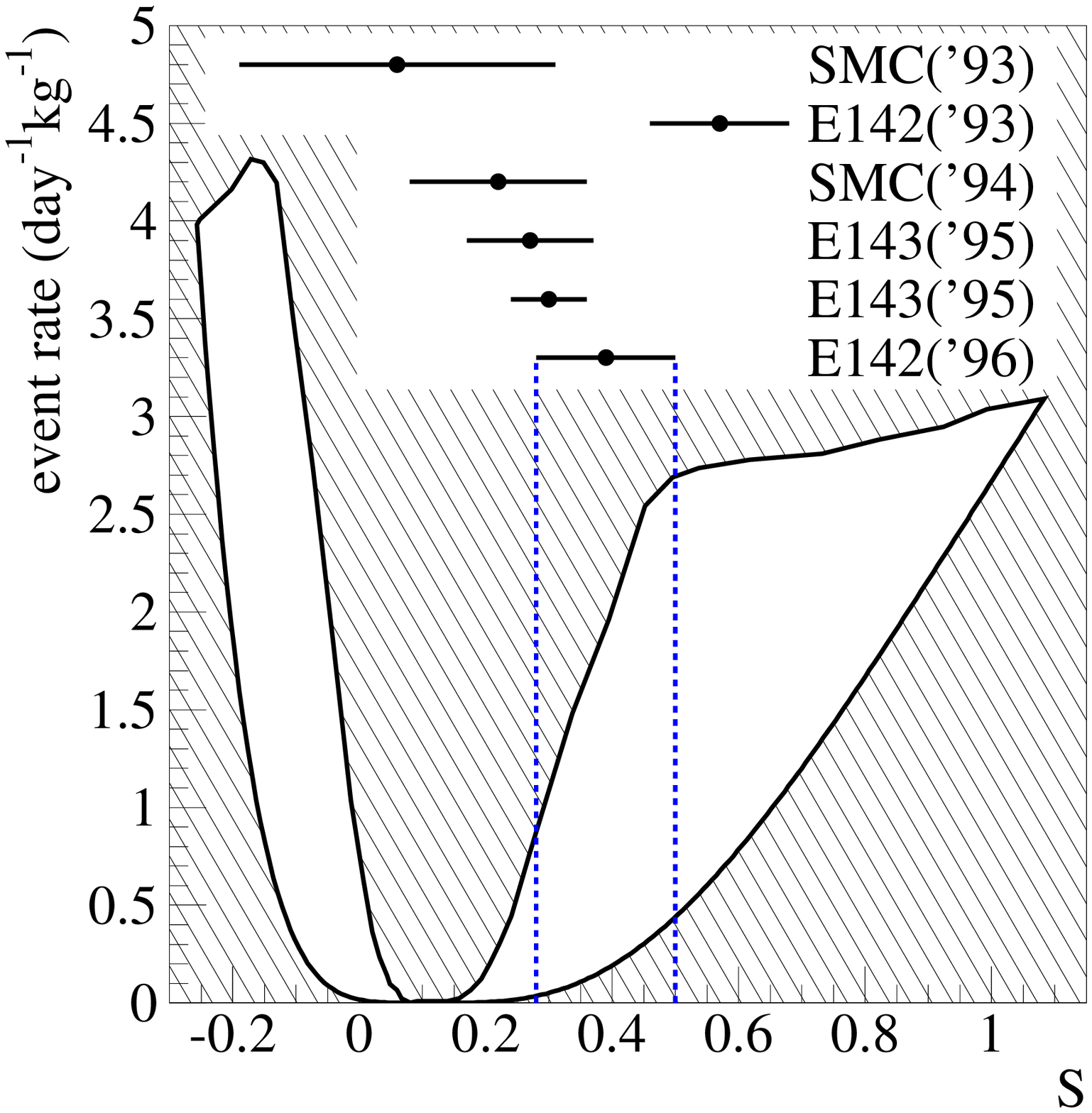}
  \end{center}
  \caption{Bound for the excitation rate.
    Hatched area is excluded by the argument of
    red-giants evolution and SN1987A \protect\cite{Haxton91}.
    Also shown is the experimental values of $S$
    that are measured recently \protect\cite{Sref}.}
  \label{fig:limit}
\end{figure}

The event rate is calculated in Ref.\ \cite{Moriyama}.
Since it depends on a parameter $S$ which characterizes
the flavor singlet coupling, which is not accurately determined,
it is plotted as a function of $S$ in Fig.\ \ref{fig:limit}.
In the figure, the experimental values of $S$
obtained recently \cite{Sref} are included.
The most recent experimental result gives $S=0.39\pm0.11$
and discussions in Ref.\ \cite{Newanalysis} indicate $S\sim 0.5$.
If the true value of $S$ is around 0.4 as suggested,
the expected event rate is restricted between 0.1 and 2.5 events
kg$^{-1}$day$^{-1}$.
Since there are both the upper bound and lower bound of
the expected event rate, experiments which have
adequate sensitivity will definitely
determine whether the axion exists with $f_a$
in the hadronic axion window.

\section{Experimental method}
We now turn to a discussion of experimental realities.
After the excitation of the nuclei by the axion,
there are two possible ways to relax.
One is the emission of a $\gamma$ ray with an energy of 14.4\,keV.
The other is the emission of an internal conversion electron
with an energy of 7.3\,keV and atmic radiations.
Since the attenuation length of the $\gamma$ ray is $20\,\mu$m
and the range of the electron is $0.2\,\mu$m in iron,
it is difficult to detect the $\gamma$ rays
or electrons outside the iron. 
In addition, detectors should have low energy threshold,
low background, and high energy resolution.
However, these difficulties are possibly overcome
by using a bolometric technique
with an absorber which contains $^{57}$Fe-enriched iron.
The technique has much advantages
compared with other techniques in these respects.
It is generally accepted that
a sensitivity down to 0.1 events kg$^{-1}$day$^{-1}$
is reachable with a bolometer for dark matter search.
If we can utilize this technique for the proposed experiment,
it is possible to obtain a definite result as discussed above.

Recently, an experimental result based on this proposal was reported.
Kr\v cmar \etal\ \cite{Krcmar} used a simple method
to detect $^{57}$Fe solar axions.
They put a thin disk made of $^{57}$Fe
in front of a silicon lithium detector.
A disk of natural iron with the same dimensions was used
for background measurement.
In this situation, they can detect only gamma rays
which escape from the target.
Since this is very rare, the total efficiency is extremely low.
Their limit is that the axion mass is smaller than 745\,eV.
Although this is far beyond the hadronic axion window,
I think this is a very interesting result.

\section{Summary}
In summary, a new scheme to detect almost monochromatic
solar axions using resonant excitation of $^{57}$Fe is introduced.
$^{57}$Fe is rich in the Sun and its first excitation energy
is low enough to be excited thermally.
Therefore, one can expect
the nuclear deexcitation accompanied with the axion emission.
Because of the Doppler effect associated 
with the thermal motion of $^{57}$Fe in the Sun,
a small portion of the axions from the nuclide
can be absorbed by the same kind of nuclide in a laboratory.
The nuclide is considered as a high sensitive detector of the axions.
The excitation rate is expected to be order of $1\,\rm day^{-1}kg^{-1}$.
Although it is difficult to detect the excitation outside the iron,
it is detected efficiently by a bolometric technique
with an absorber which contains $^{57}$Fe-enriched iron.
I am planning an experiment to search for
the monochromatic axions from the Sun in this new scheme.

\section*{Acknowledgements}
I want to thank Dr. S.~Chang
for his kindness and for discussion with him.

\end{document}